\pdfoutput=1

\documentclass[preprint,5p,numbers,sort&compress]{elsarticle}

\usepackage{graphicx}
\usepackage{amssymb,amsmath}
\usepackage{natbib}
\usepackage{lipsum}
\usepackage{hyperref}
\hypersetup{pdftitle = The title of my PDF, pdfauthor = My name, pdfsubject= The subject, pdfkeywords = keyword1 keyword2 keyword3}
\hypersetup{colorlinks = true, linkcolor = blue, anchorcolor = red, citecolor = blue, filecolor = red, pagecolor = red, urlcolor = blue}

\begin{document}

\begin{frontmatter}

\title{Fractal basins of convergence of a seventh-order generalized H\'{e}non-Heiles potential}

\author[eez]{Euaggelos E. Zotos\corref{cor1}}
\ead{evzotos@physics.auth.gr}

\author[fld]{Fredy L. Dubeibe}

\author[fld]{ A. Ria\~no-Doncel}

\cortext[cor1]{Corresponding author}

\address[eez]{Department of Physics, School of Science,
Aristotle University of Thessaloniki,
GR-541 24, Thessaloniki, Greece}

\address[fld]{Facultad de Ciencias Humanas y de la Educaci\'on, 
Universidad de los Llanos, 
Villavicencio, Colombia}

\begin{abstract}
This article aims to investigate the points of equilibrium and the associated convergence basins in a seventh-order generalized H\'{e}non-Heiles potential. Using the well-known Newton-Raphson iterator we numerically locate the position of the points of equilibrium, while we also obtain their linear stability. Furthermore, we demonstrate how the two variable parameters, entering the generalized H\'{e}non-Heiles potential, affect the convergence dynamics of the system as well as the fractal degree of the basin diagrams. The fractal degree is derived by computing the (boundary) basin entropy as well as the uncertainty dimension.
\end{abstract}


\end{frontmatter}

\section{Introduction}
\label{intro}

It is well known that every differentiable symmetry of the action of a physical system has a corresponding conservation law. Therefore, by Noether's theorem in every stationary axisymmetric system, the energy and the angular momentum along the symmetry axis are conserved. However, at the end of the XIX century it was shown that in some cases there exists an additional hidden conserved quantity (see e.g. \cite{S90,S93}), the so-called third integral of motion. This discovery increased the interest of researchers who initiated systematic studies in this topic, among whom Contopoulos stands out by his studies on the existence of the third integral of motion in galactic dynamics  \cite{C57,C60,CB62,C63A, C63B}. 

An important landmark on the existence of the third integral of motion in axisymmetric potentials is provided by the work of Michael H\'enon and Carl Heiles \cite{HH64}, who performed a systematic and complete numerical investigation on this topic, finding that the third integral exists for only a limited range of initial conditions. The potential selected for the study in Ref. \cite{HH64}, can be considered a particular case of the general Hamiltonian found by Contopoulos in \cite{C57},   
\begin{equation}\label{hc}
H=\frac{1}{2}\left(\dot{x}+\dot{y}+\omega_{1}^{2} x^2+\omega_{2}^2 y^2 +\epsilon x y^2 +\epsilon' x^3\right).
\end{equation}
setting $\omega_{1}=\omega_{2}=\epsilon=1$ and $\epsilon'=-1/3$, and by swapping variables $(x,y)\rightarrow (y,x)$.

The Hamiltonian presented above (\ref{hc}) (and consequently the H\'enon-Heiles potential) can be derived as a series expansion up to the third-order of the effective potential for stationary axisymmetric systems with reflection symmetry $V(r,z)=V(r,-z)$ (see \cite{CB02})
\begin{equation}\label{pote}
V(r,z)=U(r,z)+L^{2}/2 r^{2}.
\end{equation}
Since then, some efforts have been made to generalize the Henon-Heiles potential. Around 1980, Verhulst  \cite{V79} expanded the potential (\ref{pote}) up to the fourth-order seeking to study resonances 1:1, 1:2, 1:3 and 2:1. Some years ago, a generalized H\'enon-Heiles potential was derived by expanding up to the fifth-order the effective potential, aiming to study the equilibrium points and basins of convergence of the new potential \cite{ZRD18} and to analyze the dynamical effect on bounded and unbounded orbits of including higher-order terms in the series expansion \cite{DRZ18}. More recently, a seventh-order version of the stationary axisymmetric potential was presented \cite{DZC20}, finding that when higher-order contributions of the potential are taken into account, the chaoticity of the system is reduced in comparison with the lower-order version of the H\'enon-Heiles system.

The practical importance of the Henon-Heiles like potentials lies in its applications to the stellar kinematics and velocity ellipsoid in our galaxy, where the observed distribution of star's velocities near the Sun can be explained if a third integral exists \cite{CR20}. Also, these potentials have been used to investigate quantum manifestations of chaos and level repulsion in classical chaotic Hamiltonians \cite{CG89}, and to calculate the lifetimes and energies for metastable states exploiting the property that the dynamics of this potential changes from quasiperiodic to chaotic for higher energies \cite{WM81}. In the context of general relativity, these potentials have been used to analyze the emission of gravitational waves and to show the differences among wave emissions from regular and chaotic motion\cite{K98}, to study the geodesic motion of test particles in vacuum gravitational pp-wave spacetimes \cite{VP00}, or to perform numerical investigations related to the integrability of orbits of test particles moving around a black hole representing the galactic center \cite{VL96}, just to name some examples.

In this paper, we rewrite the general form of the seventh-order potential \cite{DZC20} in terms of two arbitrary parameters $\alpha$  and $\delta$ denoting the contributions of the fifth and seventh-order terms, in which the constants are set in such a form that the new potential exhibits an increasing number of fixed points for some values of the free parameters\footnote{Note that in Ref. \cite{DZC20} the number of fixed points is always four.}. Aiming to perform a full numerical analysis of the new potential, we shall investigate the existence of equilibrium points using the standard Newton-Raphson iterative scheme. In particular, we will use the so-called basins of convergence \cite{NY96} in order to explore the optimal initial conditions for which the numerical method is faster and accurate (see e.g. \cite{D10,KGK12,Z16,Z17}). Moreover, using the probability density function we shall analyze the influence of the free parameters on the convergence of the Newton-Raphson scheme. The fractal degree of the basin diagram will be investigated through the basin entropy and the boundary basin entropy introduced recently by Daza et. al \cite{DWGGS16, DGGWS17, DWGGS18}.

The present paper is organized as follows: In section \ref{mod}, the derivation of the generalized potential along with the new approximate potential is presented. Applying the standard linear stability analysis, in section in \ref{libs} the existence and stability of the libration points of the system are calculated as a function of two parameters $\alpha$ and $\delta$ related to the contribution of higher-order terms. In section \ref{NR}, the Newton-Raphson basins of convergence are presented using color code diagrams. Also, we show the biparametric evolution of the basin entropy, the boundary basin entropy and the uncertainty dimension as a function of $\alpha$ and $\delta$. Finally, in section \ref{conc} we present the main conclusions of our numerical study.

\section{The model potential}
\label{mod}

As already pointed out in the introduction section, in a previous paper \cite{DZC20} we derived a generalization of the H\'{e}non-Heiles potential through a Taylor series expansion up to the seventh-order of a generic potential with axial and reflection symmetries. The effective potential is of the form $V(r,z)=U(r,z)+L^{2}/2 r^{2}$, where $r$ and $z$ denote the radial distance and height of the usual cylindrical coordinates, with $V(r,z)=V(r,-z)$. The seventh-order approximate potential can be written as

\begin{eqnarray}\label{eq:2.1}
V(\xi,z)&\approx&a_{1} \xi^4+z^4 \left(a_{2}+b_{2} \xi +c_{3} \xi ^2+d_{4} \xi^3\right)+z^2 \nonumber\\
&\times&\left(a_{3} \xi ^2+b_{3} \xi ^3+c_{4}
\xi^4+d_{3} \xi^5+\omega_{2}^2+\xi \epsilon \right)\nonumber\\
&+&\beta \xi^3+b_{1} \xi^5+c_{1} \xi^6+z^6 (c_{2}+d_{2} \xi)\nonumber\\
&+&d_{1} \xi^7+\xi^2 \omega_{1}^2,
\end{eqnarray}
with 
\begin{eqnarray}\label{eq:2.2}
&&\xi=r-r_{0}, \nonumber\\
&&
\omega_{1}^{2}=\frac{3 L_z^2}{2 r_0^4}+\left. \frac{1}{2}\frac{\partial^{2} V_{\rm eff}}{\partial r^2}\right\vert_{*},  \omega_{2}^{2}=\left. \frac{1}{2}\frac{\partial^{2} V_{\rm eff}}{\partial z^2}\right\vert_{*},   \nonumber\\
&&
\epsilon= -\left. \frac{1}{2}\frac{\partial^{3} V_{\rm eff}}{\partial r \partial z^2}\right\vert_{*}, 
\beta=-\frac{2L_z^{2}}{r_0^5}+\left. \frac{1}{6}\frac{\partial^{3} V_{\rm eff}}{\partial r^3}\right\vert_{*},  \nonumber\\
&&
a_{1}= \frac{5L_z^{2}}{2 r_0^6}+\left. \frac{1}{24}\frac{\partial^{4} V_{\rm eff}}{\partial r^4}\right\vert_{*}, 
a_{2}=\left. \frac{1}{24}\frac{\partial^{4} V_{\rm eff}}{\partial z^4}\right\vert_{*},  \nonumber\\
&&
a_{3}= \left. \frac{1}{4}\frac{\partial^{4} V_{\rm eff}}{\partial r^2\partial z^2}\right\vert_{*}, 
b_{1}=-\frac{3L_z^{2}}{r_0^7}+\left. \frac{1}{120}\frac{\partial^{5} V_{\rm eff}}{\partial r^5}\right\vert_{*},  \nonumber\\
&&
b_{2}=\left. \frac{1}{24}\frac{\partial^{5} V_{\rm eff}}{\partial r \partial z^4}\right\vert_{*}, 
b_{3}=\left. \frac{1}{12}\frac{\partial^{5} V_{\rm eff}}{\partial r^3 \partial z^2}\right\vert_{*}.
 \nonumber\\
&&
c_{1}=\left. \frac{7L_z^{2}}{12 r_0^8}+\frac{1}{720}\frac{\partial^{6} V_{\rm eff}}{\partial r^6}\right\vert_{*}, 
c_{2}=\left. \frac{1}{720}\frac{\partial^{6} V_{\rm eff}}{\partial z^6}\right\vert_{*},
 \nonumber\\
&&
c_{3}=\left. \frac{1}{48}\frac{\partial^{6} V_{\rm eff}}{\partial r^2 \partial z^4}\right\vert_{*}, 
c_{4}=\left. \frac{1}{48}\frac{\partial^{6} V_{\rm eff}}{\partial r^4 \partial z^2}\right\vert_{*},
 \nonumber\\
&&
d_{1}=\left. -\frac{4L_z^{2}}{r_0^9} + \frac{1}{5040}\frac{\partial^{7} V_{\rm eff}}{\partial r^7}\right\vert_{*},
d_{2}=\left. \frac{1}{720}\frac{\partial^{7} V_{\rm eff}}{\partial r \partial z^6}\right\vert_{*},
 \nonumber\\
&&
d_{3}=\left. \frac{1}{240}\frac{\partial^{7} V_{\rm eff}}{\partial r^5 \partial z^2}\right\vert_{*},
d_{4}=\left. \frac{1}{144}\frac{\partial^{7} V_{\rm eff}}{\partial r^3 \partial z^4}\right\vert_{*},
\end{eqnarray}

where $\left.\right\vert_{*}$ denotes evaluation at $(r_0, 0)$. 

It should be pointed out that unlike our previous study, here we redefine the constant factors of the polynomial in order to obtain a large spectrum of fixed points. Also, we introduce two arbitrary parameters $\alpha$ and $\beta$, such that setting $\alpha=\delta = 0$ the new potential reduces to the well-known classical H\'enon-Heiles potential. The specific replacements are as follows: $z\rightarrow x$, $\xi\rightarrow y$,
$a_{1}=a_{2}=b_{1}=-b_{2}=-b_{3}=-\delta$, $a_{3}=-2\delta$, $c_{1}=c_{2}=d_{1}=d_{2}=d_{3}=d_{4}= 2\alpha$, $c_{3}=c_{4}=\alpha$, $\omega_{1}=\omega_{2}=1/\sqrt{2}$, $\beta=-1/3$ and  $\epsilon=1$. 

Therefore, after applying the previous replacements into Eq. (\ref{eq:2.1}), the final potential reads as
\begin{eqnarray}\label{eq:2.3}
V(x,y)&=&\frac{1}{6} \left(3 x^2+3 y^2+6 x^2 y-2 y^3\right)\nonumber\\
&+&\alpha  \left[2 x^6 (y+1)+x^4 y^2 (2 y+1)+x^2 y^4\right. \nonumber\\
&\times&\left. (2 y+1)+2 y^6 (y+1)\right]+\delta \left[x^4 (y-1) \right. \nonumber\\
&+&\left.x^2 (y-2) y^2-y^4 (y+1)\right]
\end{eqnarray}

In the next sections, the main properties and characteristics of the new seventh-order potential are analyzed.

\section{Equilibrium points}
\label{libs}

\begin{figure}[!t]
\centering
\resizebox{\hsize}{!}{\includegraphics{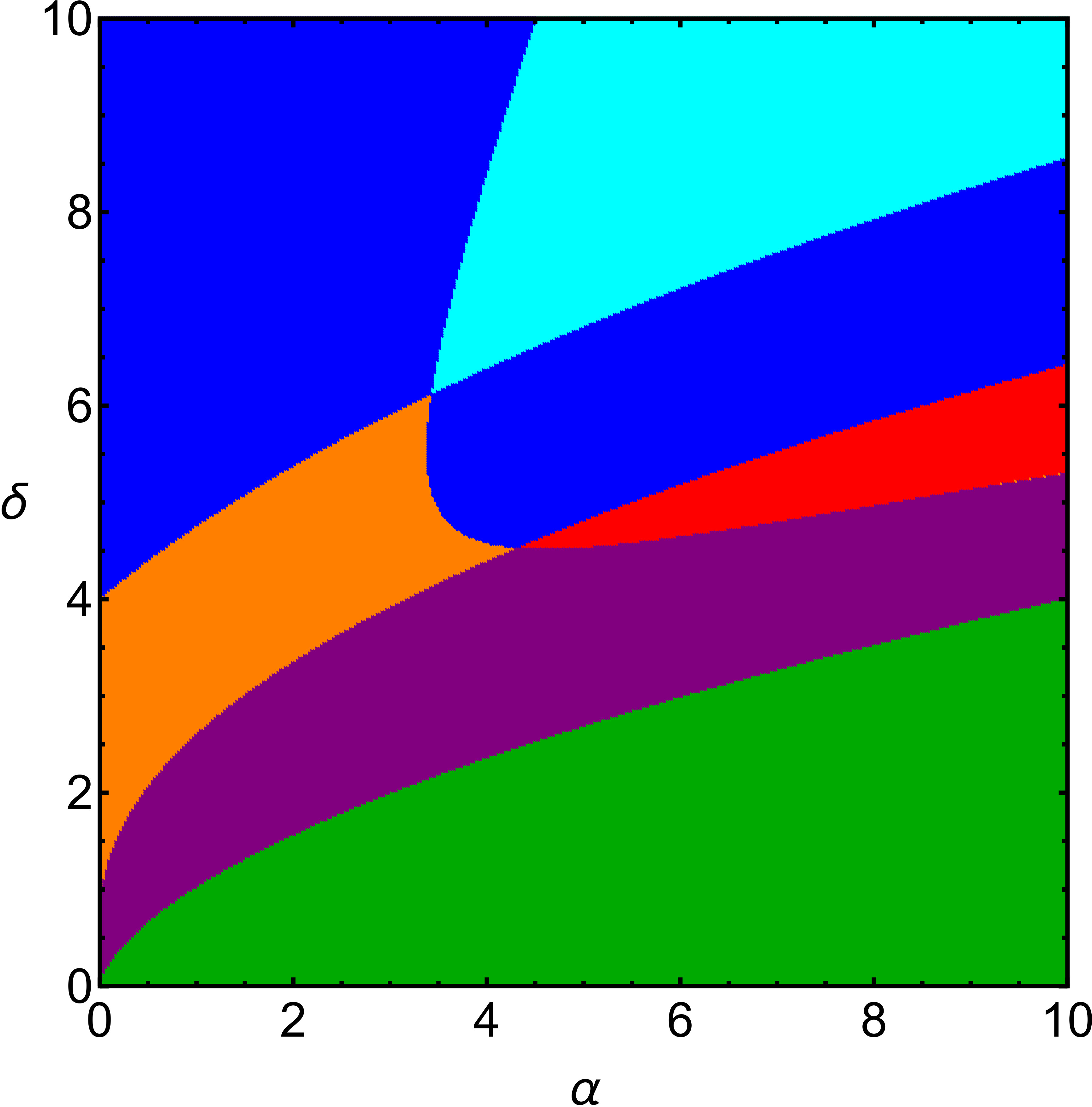}}
\caption{Color basins on the $(\alpha,\delta)$-plane, corresponding to different number of equilibrium points. 4 points (green); 6 points (purple); 8 points (red); 10 points (orange); 12 points (blue); 14 points (cyan). (Color figure online).}
\label{npts}
\end{figure}

\begin{figure*}[!t]
\centering
\resizebox{0.6\hsize}{!}{\includegraphics{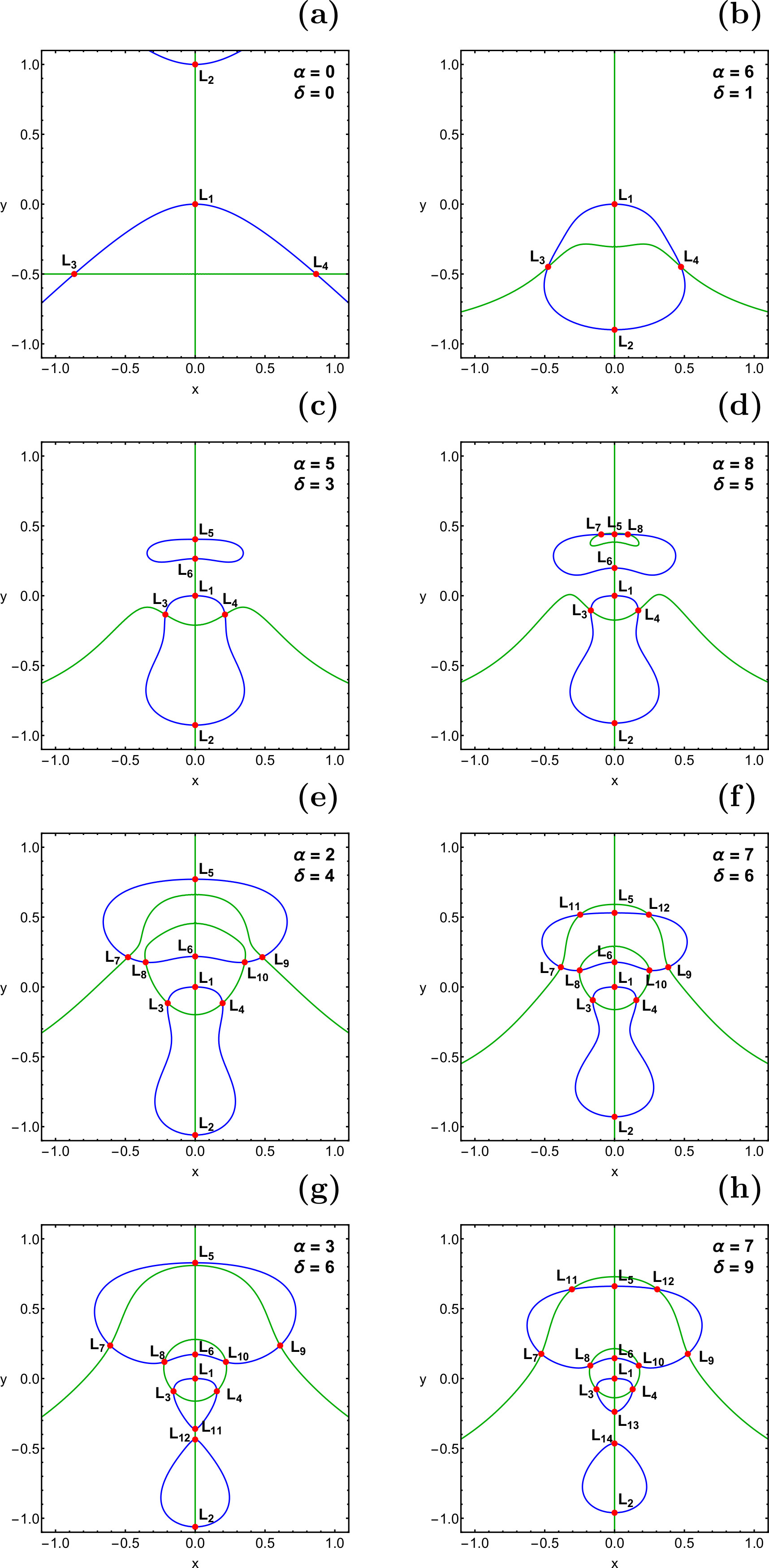}}
\caption{Contours of the equations $V_x = 0$ (green) and $V_y = 0$ (blue). The intersection points (red dots) designate the position of the equilibrium points $(L_i, \ i = 1,..., 14)$, for different values of $\alpha$ and $\delta$, corresponding to the eight different cases. (Color figure online).}
\label{conts}
\end{figure*}

The number of points of equilibrium is a function of the values of the parameters $\alpha$ and $\delta$. Our analysis suggests that when $\alpha \in [0,10]$ and $\delta \in [0,10]$ we have six cases, depending on the total number of libration points. In Fig.~\ref{npts} we present the color basins on the $(\alpha,\delta)$-plane which correspond to a different number of points of equilibrium. It is interesting to note, that in all cases the system has always an even number of libration points. Moreover, it is observed that the amount of equilibria becomes mainly affected by the two parameters $(\alpha,\delta)$ since the basins do not form vertical or horizontal bands.

Fig.~\ref{conts} shows the equilibrium positions, for eight cases, with values of $\alpha$ and $\delta$, corresponding to all possible combinations of libration points. The coordinates of the libration points are presented as the intersection points of the curves $V_x = 0$ (green lines) and $V_y = 0$ (blue lines). We should note, that in Fig.~\ref{npts} we have seen that there exist two basins corresponding to 12 points of equilibrium. It turns out that the geometry of the curves $V_x = 0$ and $V_y = 0$, as well as the locations of the equilibrium points, are different in each case. Therefore, we have eight different cases (counting also the classical HH system with $\alpha = \delta = 0$), regarding the total number of libration points.

Once the coordinates of the equilibrium conditions $(x_0, y_0)$ are determined, one can also study their linear stability. The linear stability or instability of a libration point is obtained through the following characteristic equation
\begin{equation}
\lambda^4 + \left(V_{xx} + V_{yy}\right) \lambda^2 + V_{xx} V_{yy} - V_{xy}^2 = 0,
\label{ce}
\end{equation}
where $V_{xx}$, $V_{yy}$, and $V_{xy}$ denote the second-order partial differentials of the potential $V(x,y)$ with respect to the subindex variable.

When the quartic equation (\ref{ce}) has four pure imaginary roots, then the respective point of equilibrium is linearly stable. The existence of four pure imaginary roots is secured by the three conditions
\begin{align}
V_{xx} + V_{yy} > 0&, \nonumber\\
V_{xx} V_{yy} - V_{xy}^2 > 0&, \nonumber\\
\left(V_{xx} + V_{yy}\right)^2 - 4 \left(V_{xx} V_{yy} - V_{xy}^2 \right) \geq 0&,
\end{align}
which must simultaneously be fulfilled.

Our computations indicate the following:
\begin{itemize}
  \item When 4 equilibria exist, only $L_1$ is linearly stable, while the rest of them are linearly unstable.
  \item When 6 equilibria exist, only $L_1$ and $L_5$ are linearly stable, while the rest of them are linearly unstable.
  \item When 8 equilibria exist, only $L_1$, $L_7$, and $L_8$ are linearly stable, while the rest of them are linearly unstable.
  \item When 10 equilibria exist, only $L_1$ and $L_5$ are linearly stable, while the rest of them are linearly unstable.
  \item When 12 equilibria exist (the case with the middle blue basin in Fig.~\ref{npts}), only $L_1$, $L_{11}$, and $L_{12}$ are linearly stable, while the rest of them are linearly unstable.
  \item When 12 equilibria exist (the case with the upper blue basin in Fig.~\ref{npts}), only $L_1$ and $L_5$ are linearly stable, while the rest of them are linearly unstable.
  \item When 14 equilibria exist, only $L_1$, $L_{11}$, and $L_{12}$ are linearly stable, while the rest of them are linearly unstable.
\end{itemize}
The general conclusion is that the point equilibrium located at the origin with $x = y = 0$, is always linearly stable, regardless of the particular values of the parameters $\alpha$ and $\delta$.

\section{The Newton-Raphson basins of convergence}
\label{NR}

\begin{figure*}[!t]
\centering
\resizebox{0.6\hsize}{!}{\includegraphics{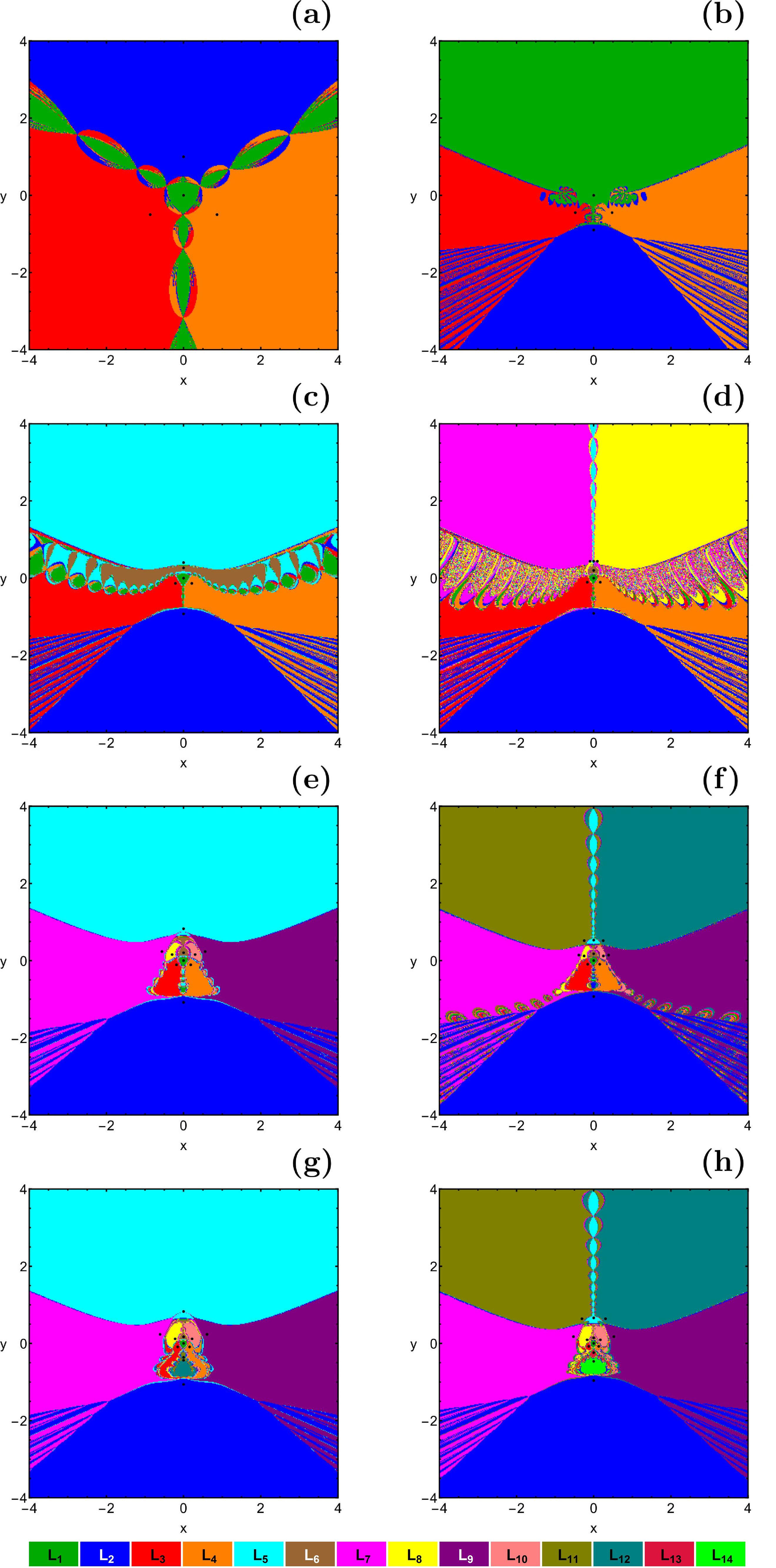}}
\caption{Basin color diagrams of the NR-BoC on the configuration $(x,y)$-plane. The values of the parameters $\alpha$ and $\delta$ are as in the respective panels of Fig.~\ref{conts}. The positions of the libration points are marked, using black dots. (Color figure online).}
\label{bas}
\end{figure*}

\begin{figure*}[!t]
\centering
\resizebox{0.6\hsize}{!}{\includegraphics{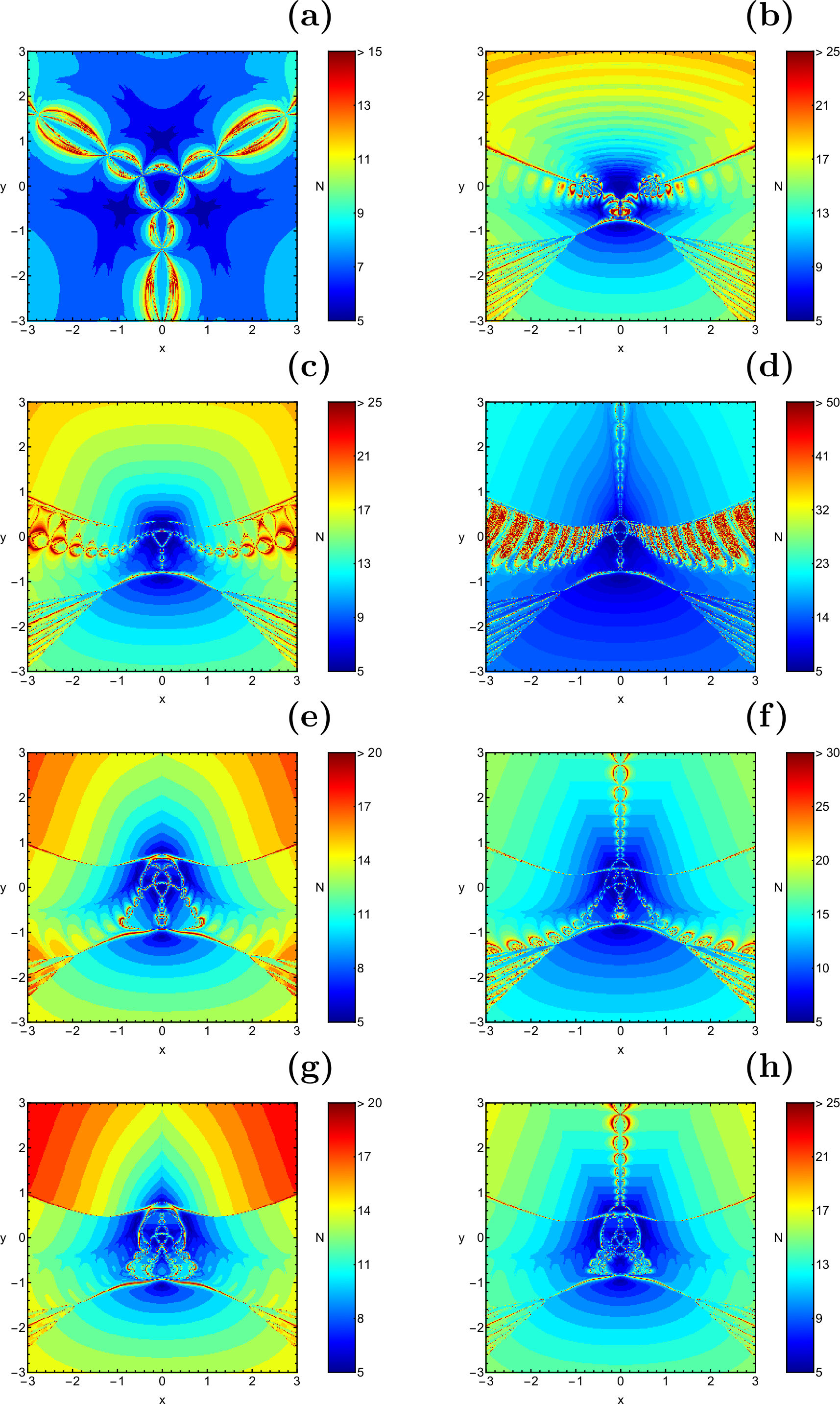}}
\caption{Color maps showing the distribution of the required number of iterations $N$, on the configuration $(x,y)$-plane. The values of the parameters $\alpha$ and $\delta$ are as in the respective panels of Fig.~\ref{conts}. (Color figure online).}
\label{iter}
\end{figure*}

\begin{figure*}[!t]
\centering
\resizebox{0.6\hsize}{!}{\includegraphics{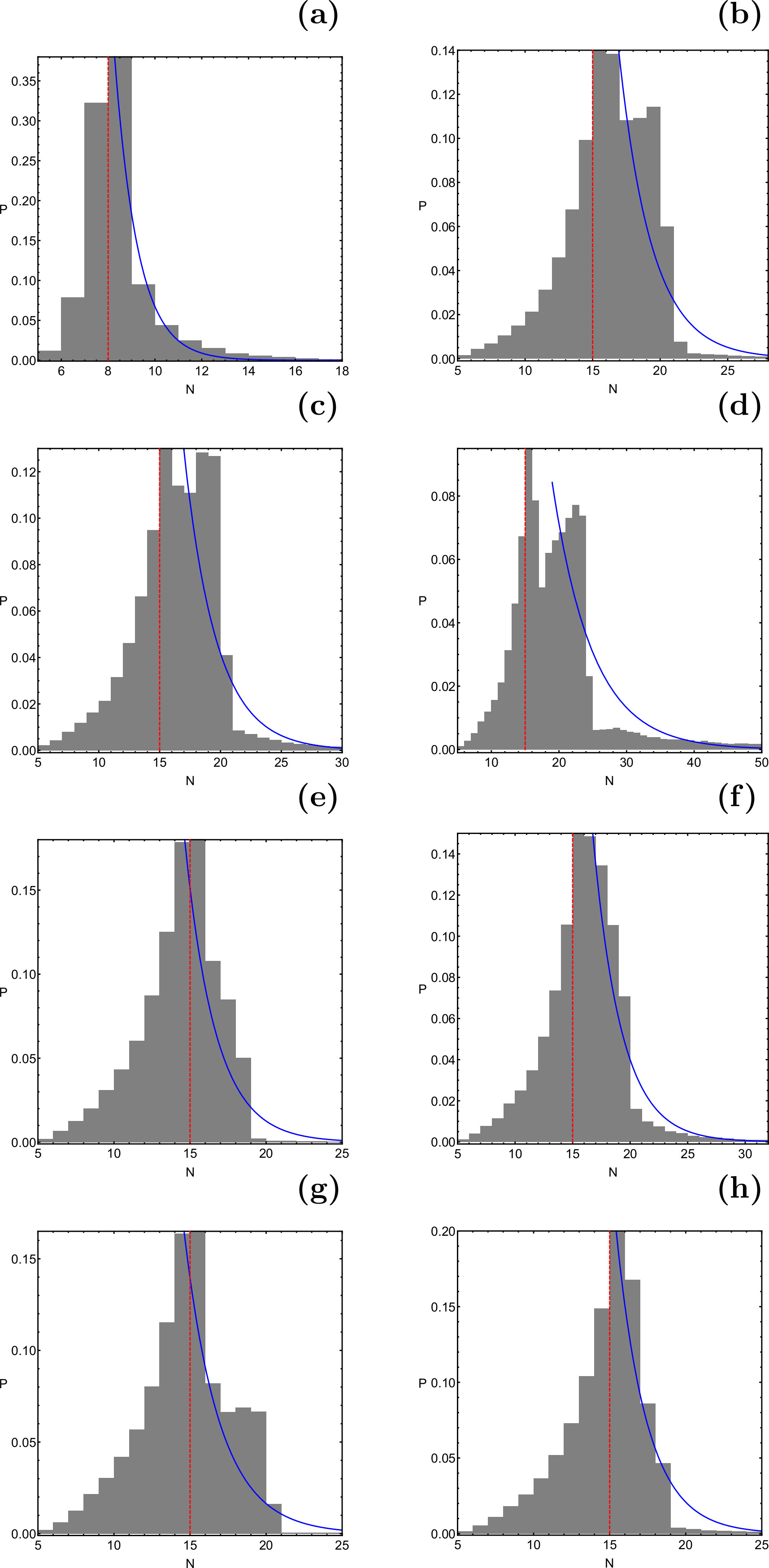}}
\caption{Probability histograms for the eight cases of Fig. \ref{bas}. The most probable number of iterations is indicated by dashed, vertical, red lines, while the blue lines correspond to the best fitting curves. (Color figure online).}
\label{prob}
\end{figure*}

Knowing the equilibrium positions of a dynamical system is very important. However, in many cases (including our modified HH system) the coordinates of the libration points cannot be derived analytically. Then, the equilibrium solutions can be derived only by employing numerical methods. One of the easiest ways of solving numerically a system of equations (in our case the coupled system $V_x = V_y = 0$) is by using the Newton-Raphson (NR) iterative scheme
\begin{align}
x_{n+1} &= x_n - \left( \frac{V_x V_{yy} - V_y V_{xy}}{V_{yy} V_{xx} - V^2_{xy}} \right)_{(x_n,y_n)}, \nonumber\\
y_{n+1} &= y_n + \left( \frac{V_x V_{yx} - V_y V_{xx}}{V_{yy} V_{xx} - V^2_{xy}} \right)_{(x_n,y_n)}.
\label{nrm}
\end{align}

It is a well-known fact, that the outcomes of any numerical method are influenced by the choice of the starting conditions. In particular, both the speed and the accuracy of any numerical scheme fully depend on the chosen initial conditions. There exist starting conditions for which the iterator diverges, while there are also exist starting conditions leading to one of the roots of the system. The ideal initial conditions (regarding fast convergence and accuracy) form the so-called NR basins of convergence (NR-BoC). This is exactly the importance of identifying the location of the NR-Boc of a dynamical system.

In panels (a)-(h) of Fig.~\ref{bas} we present the structure of the NR-Boc on the configuration $(x,y)$-plane, for the eight different cases, classified in terms of the number of equilibrium points. In all cases, the values of the parameters $\alpha$ and $\delta$ are the same as those of the panels of Fig.~\ref{conts}. For our computations, the NR scheme was allowed to perform up to 500 iterations, while the desired accuracy, regarding the $(x,y)$ equilibrium positions, was set to $10^{-16}$.

From the basin diagrams of Fig.~\ref{bas}, it is observed that many structures on the configuration $(x,y)$ plane are very intrincated. Moreover, some of the NR-BoC have a finite domain, while others extend to infinity. Nevertheless, in all cases, there exist well-defined structures containing ideal starting conditions for the numerical scheme. In Fig.~\ref{iter} we display color maps showing how the required number of iterations $N$ is distributed on the $(x,y)$-plane. Furthermore, in Fig.~\ref{prob} we provide the probability distributions. 

The histograms displayed in Fig. \ref{prob} with the probability distributions, may provide additional information about the properties of the modified NR method. For example, the right-hand side of the histograms can be fitted by using the well-known Laplace distribution or double exponential distribution, which is the simplest and most suitable choice \cite{ML01,SASL06,SS08}.

The probability density function (PDF) for the double exponential distribution reads as
\begin{equation}
P(N | l,d) = \frac{1}{2d}
 \begin{cases}
      \exp\left(- \frac{l - N}{d} \right), & \text{if } N < l \\
      \exp\left(- \frac{N - l}{d} \right), & \text{if } N \geq l
 \end{cases},
\label{pdf}
\end{equation}
where the quantities $d > 0$ and $l$ are known as the diversity and the location parameter, respectively. Since we are interested only in the probability tails for the histograms, we need only the $N \geq l$ part of the PDF. 

\begin{figure*}[!t]
\centering
\resizebox{\hsize}{!}{\includegraphics{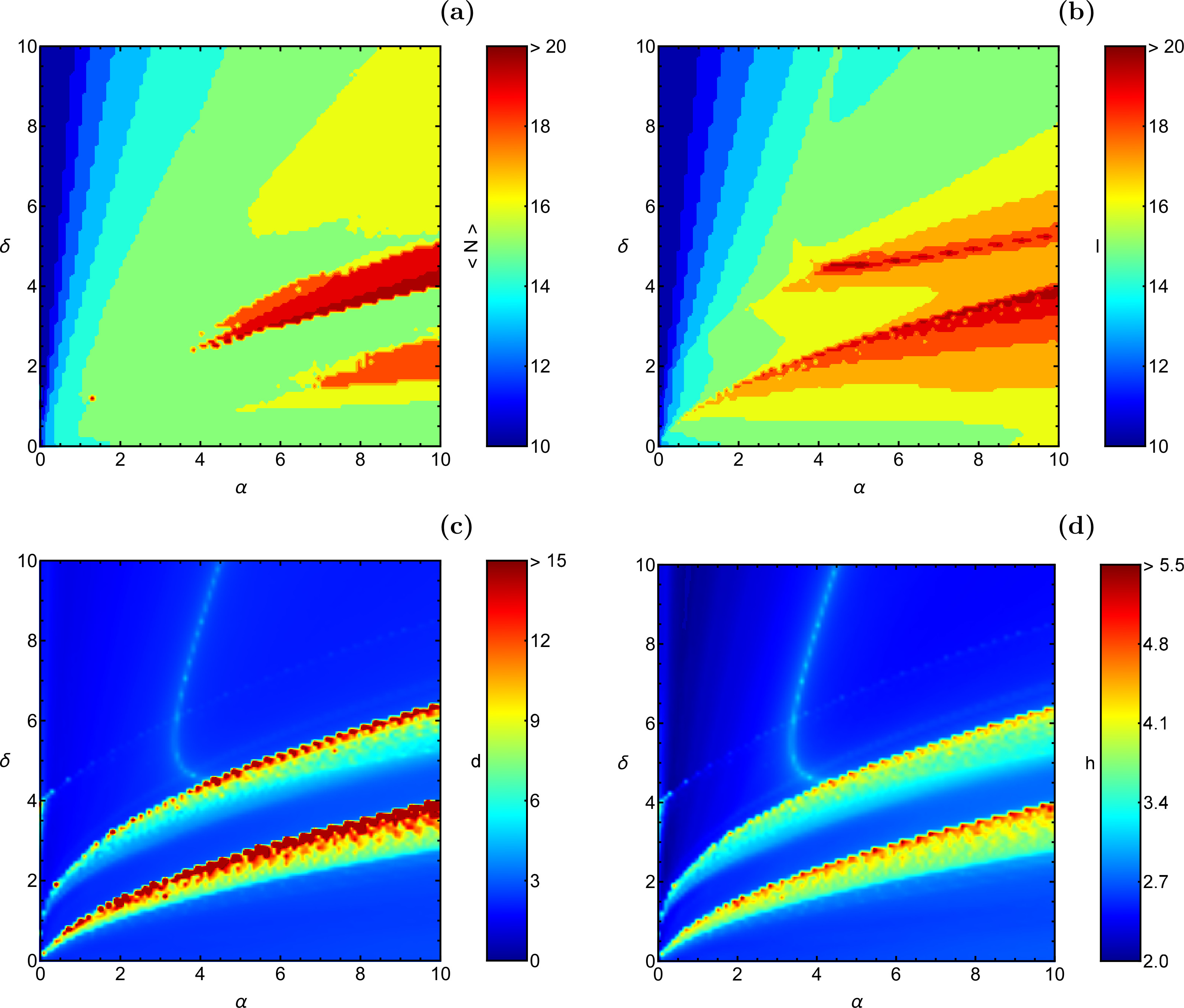}}
\caption{Biparametric evolution of (a): $< N >$; (b): $l$; (c): $d$; (d): $h$, as a function of $(\alpha, \delta)$. (Color figure online).}
\label{stats}
\end{figure*}

\begin{figure*}[!t]
\centering
\resizebox{\hsize}{!}{\includegraphics{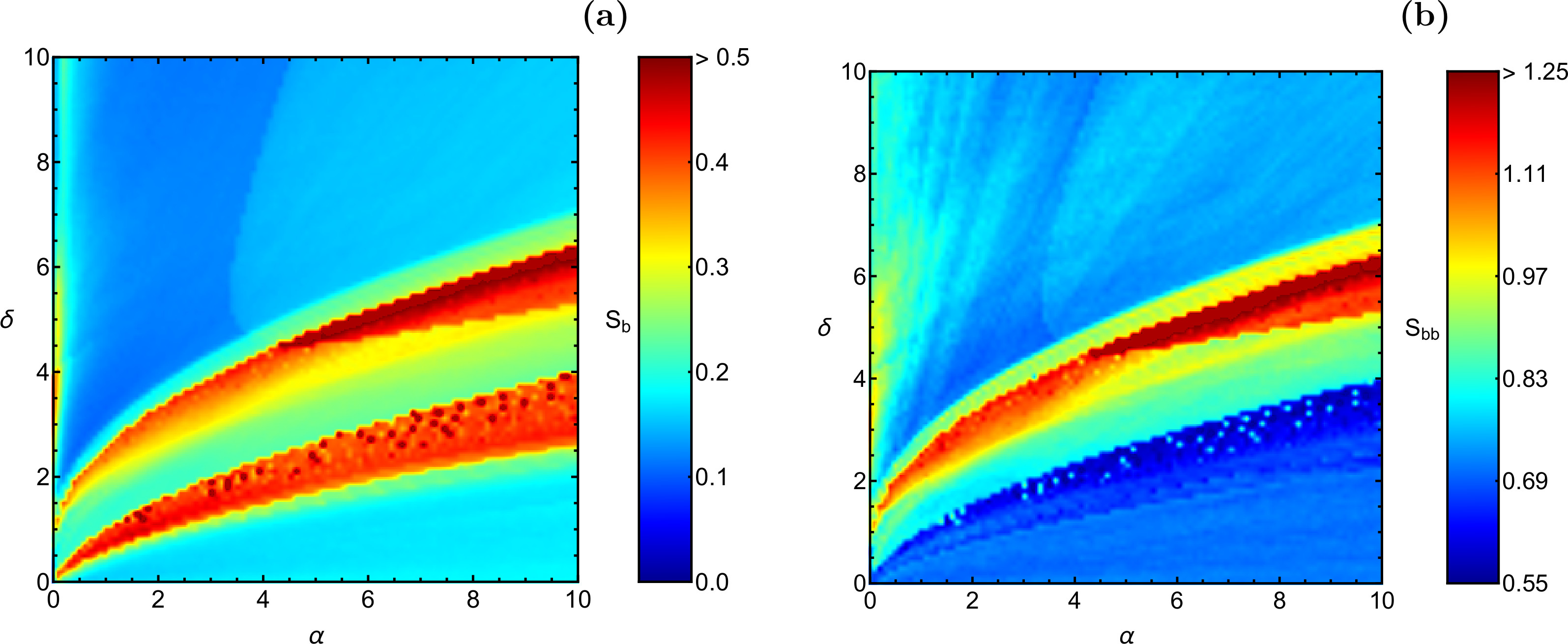}}
\caption{Biparametric evolution of the (a): basin entropy $S_b$ and (b): boundary basin entropy $S_{bb}$, as a function of $(\alpha, \delta)$. (Color figure online).}
\label{frac}
\end{figure*}

\begin{figure}[!t]
\centering
\resizebox{\hsize}{!}{\includegraphics{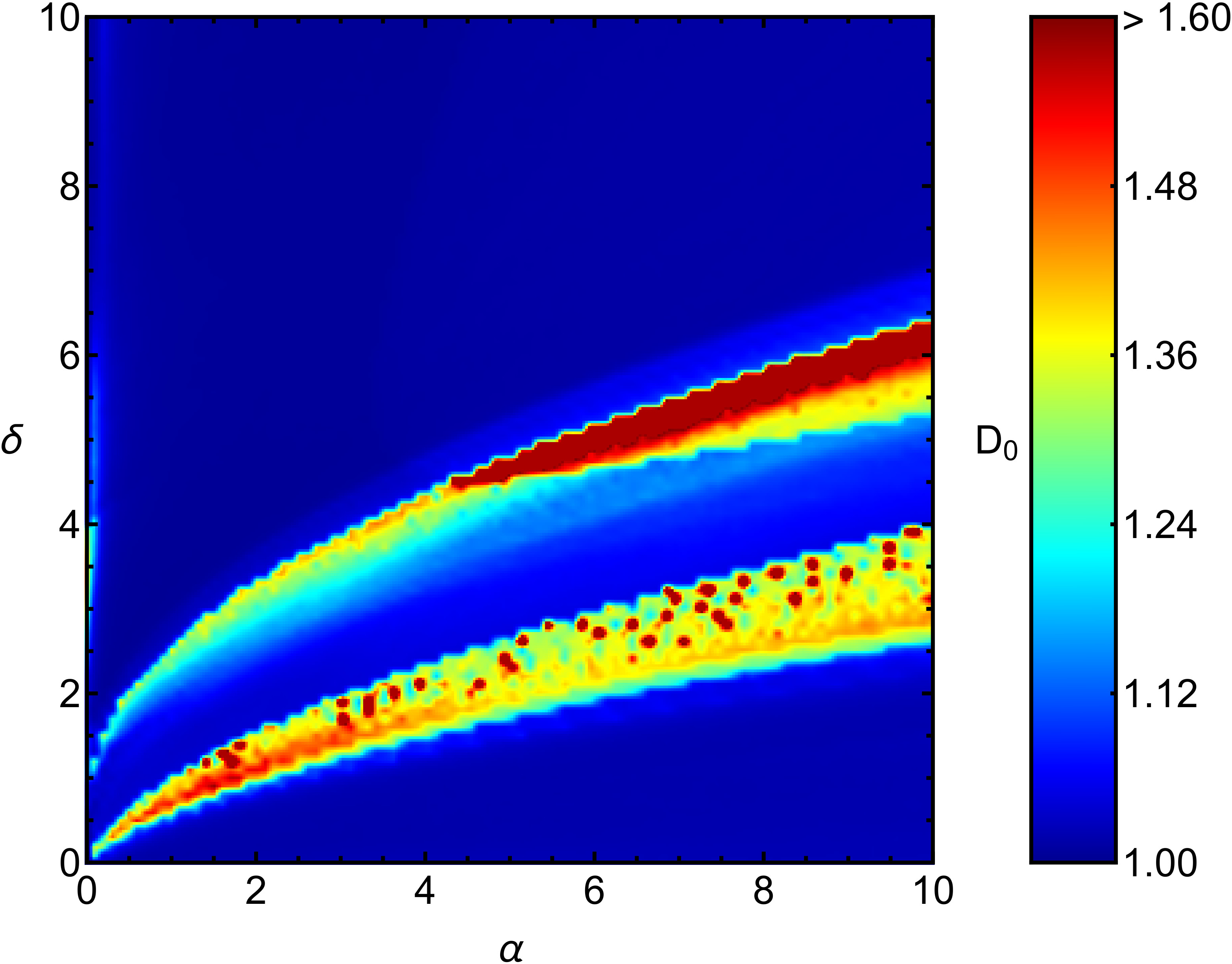}}
\caption{Biparametric evolution of the uncertainty dimension $D_0$, as a function of $(\alpha, \delta)$. (Color figure online).}
\label{d0}
\end{figure}

We aim to understand how the parameters $\alpha$ and $\delta$ influence the convergence properties of the NR scheme. To this end, we defined a $1024 \times 1024$ grid of $(\alpha,\delta)$ values and for each pair, we used the NR scheme for classifying a set of $300 \times 300$ $(x_0,y_0)$ initial conditions, on the configuration plane and in particular inside the squared region $-5 \leq x,y \leq +5$.

In part (a) of Fig.~\ref{stats}, we present the evolution of the average number of iterations $\langle N \rangle$, needed by the NR method for providing the coordinates of the equilibria with the desired accuracy. Panels (b) and (c) of Fig.~\ref{stats} depict the distributions of the location parameter $(l)$ and the diversity $(d)$ of the Laplace PDF. Our results strongly indicate that the Laplace PDF is an excellent candidate for fitting the probability histograms, if we take into account that the numerical values of $\langle N \rangle$ and $l$ are very close $|l - \langle N \rangle| \leq 2$). Additionally, from the distribution of the diversity $d$, shown in part (d), we can conclude that the probability histograms are very well-organized around the average value $\langle N \rangle$, since in most of the cases the numerical value of the diversity is relatively low $(d < 5)$. Finally, in panel (d) of Fig.~\ref{stats}, we show how the differential entropy, defined as $h = 1 + \ln(2d)$, evolves as a function of the values $(\alpha, \delta)$. It is seen, that both quantities $d$ and $h$ have a very similar parametric evolution. If we take into consideration the combined information from all four panels of Fig.~\ref{stats} we can argue that the NR method works faster when the system has either 4, 10, 12 or 14 points of equilibrium, while when 6 or 8 libration points exist the convergence of the NR scheme is considerably slower.

Previously, in Fig.~\ref{bas} we have seen that there are certain regions on the plane $(x,y)$, where using the corresponding starting conditions it is very difficult to know beforehand to which point of equilibrium they are going to converge. These regions are composed of a fractal mixture of final states (equilibria) and they are of course the exact opposite of the basins of convergence. In order to obtain quantitative information about the fractal degree of the BCs on the plane $(x,y)$, we shall compute the basin entropy $S_b$ \cite{DWGGS16,DWGGS18}. This modern tool indicates the fractal degree of a basin diagram by examining its topological properties. In part (a) of Fig.~\ref{frac} we show the distribution of the numerical values of $S_b$, as a function of $(\alpha, \delta)$. Now we can conclude, without any doubt, that when the system has eight points of equilibrium, we encounter the most fractal NR-BoC, while the fractal degree is considerably lower for a higher number of libration points.

Unfortunately, the transition between smooth and fractal boundaries cannot be determined by the basin entropy $S_b$. The main reason for this drawback is that the basin entropy addresses the uncertainty to link a set of initial conditions to its corresponding final states. Therefore, if we are interested in detecting small variations in the basin boundary we must use another indicator, the boundary basin entropy $S_{bb}$, which was introduced for the first time in 2016 by Daza et.al. \cite{DWGGS16}. For obtaining the boundary basin entropy, all we have to do is to divide the total entropy between the number of cells that fall in the boundaries of the convergence basins. This tool gives us the possibility to safely conclude if the basin boundary is fractal or not, by using the so-called ``log 2 criterion", with the sufficient condition, if $S_{bb} > \ln 2$, then the boundary is certainly fractal. The distribution of the values of $S_{bb}$, as a function of $(\alpha, \delta)$, is given in panel (b) of Fig.~\ref{frac}. We see that when eight points of equilibrium exist the basin boundaries on the $(x,y)$-plane are always fractal, while on the other hand when the system has only 4 libration points, the basin boundary entropy exhibits the smaller values when compared to the other cases.

Finally, another standard way to measure the level of fractality of a basin diagram is by computing the fractal dimension \cite{O93}. At this point, it is important to emphasize that the results obtained with the basin boundary entropy $S_{bb}$ and the fractal dimension $D_{0}$ are related but they do not necessarily have to be the same because the first numerical tool allows us to assess easily that some boundaries are fractal, while the second one provides information about the whole basin since the fractal dimension is an intrinsic property of the system \cite{AVS01,AVS09}. In Fig. \ref{d0}, we present the dependence of the uncertainty dimension $D_{0}$ with the parameters $\alpha$ and $\delta$. As usual, when the fractal dimension equals one, the fractality is zero, while if its value tends to 2 it suggests complete fractality of the respective basin diagram. It is seen, that $D_0$ displays the highest values when eight points of equilibrium exist, while the lowest values are observed for the cases with 10, 12 and 14 libration points. One should certainly note the large similarity on the parametric evolutionary pattern of $D_{0}$ with respect to that of the basin entropy $S_b$. This similarity can be explained by considering that these two computer-based analysis techniques are grounded on box-counting methodologies.

\section{Discussion}
\label{conc}

In this work we explored, using numerical techniques, the equilibrium points and the convergence properties of the associated basins of convergence, of a seventh-order generalized H\'{e}non-Heiles potential. The Newton-Raphson root method was used for locating the $(x,y)$ coordinates of the points of equilibrium, while their linear stability was also revealed as a function of both parameters $\alpha$ and $\delta$. Modern color-coded plots were deployed for illustrating the convergence basins on the $(x,y)$ plane. Finally, we managed to determine how the parameters $\alpha$ and $\delta$ affect both the accuracy and speed of the NR method, while the fractal degree of the respective basin diagrams was estimated by computing the (boundary) basin entropy and the uncertainty dimension.

The routine of the bivariate NR scheme was coded in \verb!FORTRAN 77! (see e.g., \cite{PTVF92}). For the taxonomy of the starting points on the plane $(x,y)$ we needed, per grid, roughly about 3 minutes using a Quad-Core i7 4.0 GHz CPU. All the plots of the paper have been developed by using the software Mathematica$^{\circledR}$ \cite{W03}.

\section*{Acknowledgments}

This work was partially supported by  COLCIENCIAS (Colombia) Grant 8863 and by Universidad de los Llanos.

\section*{Compliance with Ethical Standards}
\footnotesize

The authors declare that they have no conflict of interest.

\end{document}